\documentclass{article}
\usepackage{graphicx} 

\title{LIMITS AND EPISTEMOLOGICAL BARRIERS TO THE HUMAN KNOWLEDGE OF THE NATURAL WORLD}
\author{J.E. Horvath, R. Rosas Fernandes and T. P. Idiart} 
\date{IAG-USP, S\~ao Paulo, Brazil\\foton@iag.usp.br}

\begin{document}

\maketitle
ABSTRACT: The goal of this article is to give an overview of  the current limitations and epistemological barriers in Science 
and Scientific Philosophy from a very general point of view. We first list and define the types of knowledge nous, doxa and episteme, 
and the Sobject-Observer and Object(s) of study, to proceed showing the different types of barriers that difficult the knowledge of the 
physical world: limitations in the language, in the logic of the Subject-Observer. Later, we discriminate between technological barriers, 
(temporary) limits and absolute epistemic barriers. The last type of limits are presented and discussed in some detail: the quantum of action, 
Planck’s scale and quantum gravity (showing the importance of the trans-Planckian scale for structure formation), the cosmological horizon 
(a limit to the present observable Universe) and the event horizons (disconnecting the inside of some spacetimes from the rest of the Universe).  
We argue that physical problems in which absolute barriers seem to determine the end of the attainable knowledge, are in fact amenable to be studied, at least indirectly.
\bigskip


\vfill\eject
\section{Introduction}

The purpose of this article is to bring to light the limits and epistemological barriers that exist in Science and Scientific Philosophy. 
Starting from the examination of the relationship 
between the Subject-Observer (denomination adopted in this work, patterned after Quantum Mechanics) and the Object-Phenomenon, this 
article will highlight the limits and barriers existing in each of them, and those existing in the means used in the search for knowledge, 
that is, the technology used for this purpose. That said, two basic differentiations must be made: epistemological limits are understood 
as an impediment that partially or temporarily limits scientifically based knowledge and, by epistemic barrier we mean an insurmountable 
and definitive impediment in the search for scientific knowledge, if such a thing exists at all.

\subsection{Nous, doxa and episteme}

The way we know the world has a long and venerable history. It was up to the pre-Socratic philosopher Anaxagoras of Miletus to present 
the concept of nous, that is, of absolute knowledge, superior to any other, attributed by him to a Cosmic Intellect. This idea of superior 
knowledge was adopted and modified later. For example, Plato uses this concept in several ways, mainly associated with consciousness and 
Aristotle identifies nous with the Prime Mover. For several centuries early in the history of Western thought, nous was the highest 
objective to be achieved by knowledge, going beyond mere logos. For Anaxagoras, knowledge via logos would lead to episteme only, 
that is, to a rational description and understanding of the world. To achieve nous, absolute and profound knowledge about all things that 
exist, a kind of cognitive intuition would be necessary. We know about the problems that Anaxagoras' thought raised, since it was 
appropriated and transformed not only by Plato and Aristotle, but also by mystics, gnostics and hermetics over the centuries. 
We therefore stick to follow the path proposed by Parmenides, that is, the path of logos that leads to episteme, that is, to scientific, 
objective and rationally justified knowledge, the consensual goals of modern Science.

This path proposed by Parmenides resulted from the bifurcation and distinction exposed in his famous poem, between doxa and episteme. 
Doxa, which comes from Greek, means “opinion”. Science and Philosophy work with universal and necessary questions, but not with opinions. 
The ubiquitous statement “I think...” is pure doxa. “I am of the opinion that…” is also doxa. Current Science, by construction, considers 
doxa as something irrelevant, given its subjective character, dependent on the individual's taste and, therefore, will never lead to any 
rationally justified knowledge. This was not so clear in Parmenides' time, and took a long time to be defined and established. The great 
Greek thinkers established doctrines with a large dose of doxa and mythical thinking, but progressively epistemological knowledge emerged 
and was consolidated, leaving doxa out.

A stated, currently, the episteme is the objective and core of scientific thought, rational, logical, grounded and empirically proven 
knowledge. Today we understand that scientific knowledge is never absolute, since scientific statements are at all times subject to questioning, 
but constitute the best that can be achieved given the state-of-the-art (state of knowledge) at the moment.

\subsection{Features of Epistemology}

The way we obtain knowledge from natural phenomena was, for centuries, the object of Philosophy, in the field of Theory of Knowledge, currently Science and Scientific Philosophy have adopted the issue under the name Epistemology or Epistemological Knowledge. It is worth questioning how to characterize this type of knowledge. Broadly considered, all epistemological knowledge has the following characteristics:

$\ast$	Dependent – directly linked to the information given by the object under study and the subject’s ability to understand (that is, a conceptual framework).
 
$\ast$	Evaluative – knowledge of the object cannot be subject to any type of value judgment on the part of the Subject-Observer.

$\ast$ Zetetic – knowledge describes the mute as it is or appears to be to the subject, not as the subject thinks it should be. From this perspective, all dogmatic and evaluative approaches and understandings are ruled out.
 
$\ast$ Contingent – considering the zetetic approach, knowledge is circumstantial, uncertain and unpredictable.
 
$\ast$ Systematic and/or Methodical – similarly, the approach establishes cause and effect relationships as they appear to be, and not as they should be.

$\ast$ Verifiable – the information collected by the Subject-Observer must be measurable and calculable. To this end, a specific technical language is used, as accurate and objective as possible, sometimes with an emphasis on Logic and Mathematics. Even so, sometimes, there is no way to completely get rid of any ambiguities, as, for example, in the case of Quantum Mechanics.

$\ast$ Approximately accurate – considering that the Subject is intrinsically flawed, the Object subject to change and new methodologies and technologies are created and introduced, changing information and understandings, the results will be approximately accurate, or, at least, subject to revisions.

\section{Relationship between the Subject and the Object}

For a better understanding of the characteristics exposed above, let us consider the relationship between the Subject-Observer and the Object. In this relationship, the Object presents a series of information to the Subject-Observer. Once the Subject-Observer comes to have full understanding and comprehension of this information, it can be said it became knowledge. Thus, the production of knowledge is based on the relationship between the world, where objects are found, and the mental world, where information is recorded and understood by the Subject-Observer.

Knowledge, in turn, is the way in which human beings seek to decipher how the world works. However, this relationship can present limitations and barriers on both sides, that is, both on the Object as on the part of the Subject-Observer.

\subsection{Objects}

For a scientific approach to any object, we must take into account that objects are divided into two types: Formal and Factual.

Formal objects are those which, from the point of view of knowledge, can be considered independent of their content, the matter in them or the concrete situation to which they apply – therefore of absolute accuracy: Mathematics – (which works with abstract entities) and Logic (which deals with the coherence of reasoning in the formation of arguments).

Factual objects: these objects of study refer to facts or entities actually existing in the real world and which, therefore, are dimensioned in terms of time, space, circumstances and characteristics – therefore, of variable accuracy according to their complexity.

Therefore, the Subject-Observer can place as Object of study a Formal Phenomena, arising from an abstract mental construction, as well as any other Factual Phenomenon that manifests itself in the world. It must be said that the distinction is not always clear and may be questioned, however, we will not enter in this debate. It is now time to consider the limitations and possible barriers that exist within the Subject-Observer himself.

\subsection{Limitations and barriers of the Subject-Observer}

In the search for knowledge, the Subject-Observer initially uses the five senses, each of which has its own range limitations. For example, we now know that human vision has a very small range within the spectrum of electromagnetic waves. Someone could object by claiming that technology can overcome these barriers. The answer is “yes” and “no”, as technology is just a circumstantial, temporal, changeable and compensatory aspect of the limits of our senses. Furthermore, within the ability to grasp the world, it is said that tthe subject is tied to time, space and causal relationships.

It was up to the philosopher Immanuel Kant, in his work entitled Critique of Pure Reason, to question the limits of human knowledge, independent of any sensitive experience. The conclusion that Kant (1999) reached is that the entire apprehension of human knowledge is delimited by time, space and causal relations. Circumstances and characteristics variables may be added, but these two qualities are external to the Subject's reason (Milton Greco, private communication). Thus, according to Kant, all human reasoning works within these limits: time, space and causal relationships. The propositions that Kant called synthetic a priori judgments present cognitive limitations that will become present when shaping new knowledge, generally a posteriori knowledge. However, it is important to highlight that Kant admits the existence of a priori synthetic propositions, such as those derived from Mathematics, which would allow new Formal Phenomena to be inferred without the need of empirical contents. It is also important to highlight that Kant discusses knowledge arising from the relationship between our knowable limitations, Formal Phenomena and Factual Phenomena. Thus, according to Kant, we always shape the Factual World and ignore the world as it is "in itself", or the noumena.

Add to that the methodological aspects that Descartes (1999) had presented in the Discourse on Method, namely: that the approach to the Object is carried out free from tendencies, prejudices, prejudices, personal interests or any type of bias.
However, we must go a little deeper into other issues that involve the ability to obtain information from the Object and transform it into knowledge on the part of the Subject-Observer. If reason has its own limits, so that man shapes the world by receiving information that is external to him and methods have many other limits, we can question how reason can create, develop and apply formal, abstract, perfect and exact objects, as is the case with Mathematics and Logic – after all, they are abstract languages, {\it albeit} very precise ones.

\subsection{Limitations of the language of the Subject-Observer}

Philosophy contemplates a range of possible origins for language, throughout the Greek skeptics, Rousseau, Nietzsche, until we reach Chomsky, for whom human beings have an innate capacity to develop a language, and even a universal language. Chomsky's generativism or generative theory is, therefore, another attempt to formalize linguistic facts, applying a precise, explicit and finite mathematical treatment to the properties of natural languages (Chomsky 2007).

To what extent can language express knowledge from one subject to another in an increasingly accurate and precise way? For humans, this is a difficult question to answer. Regardless of the theory adopted, language and its forms of externalization and communication are human products.

We can, therefore, question whether language has the power to transmit all our knowledge in a way free from biases. It is true that the richer and more objective the language, the greater the ability to communicate from one subject to another without errors and inaccuracies. And vice versa: there is only knowledge if it can be grasped and expressed through precise language. In this sense, language necessarily hovers over all types of knowledge. Ludwig Wittgenstein is an important figure for this problem, stating that if there are limits to knowledge, there will necessarily be limits in its expression, in language itself. Special attention is given by Wittgenstein to the meaning of language sentences, which are relevant, according to him, if they have references to the state-of-the-art of the world/state of knowledge, otherwise they lack meaning. Wittgenstein's latest works interpret words as tools for the private "games" that each individual plays, understood as everyday situations where communication is central. This set of ideas is often summarized in the phrase "the limits of language are the limits of the world", applicable at least to its initial ideas, or, alternatively, the limits of knowledge of the world are manifested in the limits of language. (Wittgenstein 2001).

However, we also think through images, sometimes difficult to verbalize, and we add other forms of communication, such as through logical, mathematical, musical symbols, among others. This is the extended meaning we must give to language. Going further, for any type of objective communication to be successful, this language must be logical, widely understandable and free of inconsistencies.

\subsection{Limitations of the logic of the Subject-Observer}

The Subject-Observer also needs a consistent logic to process the information gathered from the Object. However, this is not that simple: logicians are often divided into two basic groups, those who apply logic to the real world, and those who maintain that logic is abstract in nature and should not be applied to knowledge of the world.

We may think that the world is logical in whole or in part. If the world is fully logical as a whole, through logic, it would be possible to achieve absolute knowledge. However, this is not what happens. The world is not totally logical and unpredictable variables exist. If the world were completely logical, it would also necessarily be predictable and we would reach nous, the absolute knowledge proposed by Anaxagoras. The world is largely logical, which makes epistemological knowledge possible, as initially exposed, that is, scientific epistemological knowledge through reason (logos). On the other hand, it is possible to carry out a series of logical and/or mathematical thoughts that do not have a connection with the world as it reveals itself, or as we understand it.

Thus, if in part our language, our Logic and our Mathematics are applicable in order to understand part of the world, the opposite is also revealed: several operations are only possible within the field of Logic and Mathematics. This reveals a series of gaps between phenomena and their information, senses, reason, and understanding by the Subject-Observer. These gaps, however, are not the only limitations that can be detected.

The construction of so-called non-Aristotelian Logics (perhaps better defined as non-classical) is now a topic of great interest in Science and Scientific Philosophy, especially because they are reflected in several contemporary applications for computing and other subjects. The construction of this type of systems is possible in an analogous way to the construction of non-Euclidean Geometries: it starts by denying at least one of the ordinary postulates (for example, the one saying that a statement can only be true or false), and subsequently it is possible to construct, nevertheless, a consistent formal system. There are many different versions, depending on the postulate or denied postulates, which lead to consistent logics, in the sense that they are not self-contradictory. A particularly interesting example is the Quantum Logic of Janusz von Neumann (1932), demonstrating that a Logic that excludes the ordinary distributive property can lead to a resolution of the dilemmas of the superposition of quantum states, which are the source of paradoxes such as the one exemplified with "Schrödinger's cat" (Bunge 2012). Adopting this development leads us to admit that Nature uses a non-human type of reasoning, a consequence that is not widely accepted by scientists, but is entirely possible. We see here that the limitations of having a Boolean-Aristotelian Logic as the basis of our understanding of the physical world can be an important, if not definitive, obstacle if Quantum Logic is the way forward.

\subsection{Sociological limitations of the Subject-Observer}

We must also address psychosocial barriers. Time (or, more precisely, the modern times of Ortega y Gasset 1923) and the historical context that defines what is a scientific problem and what is Science, or what is pseudoscience and Science is in fact a psychosocial limitation to be considered. To put it more simply, each era faced different problems and issues and/or in different ways.

One type of knowledge (or rather, pseudo-knowledge) is beliefs. Acceptance of beliefs is not rational, and does not need any rigorous proof. Ergo, beliefs can be transformed by the Subject-Observer into knowledge, but the result is very fragile and uncertain knowledge. Still, some beliefs are not all bad, as they can lead us to more structured questioning, to an experiment to verify their veracity.

As a trivial example, the belief that far beyond the known land the world ended, and monsters and emptiness awaited navigators, made the Portuguese travel around the world and established once and for all that the Earth was in fact a sphere. This is obvious today, but constituted a frontier in the knowledge of the world that lasted many centuries.

Sciences feature a set of extremely persistent beliefs, for example, the belief that the Universe was perfect, that the stars were fixed and that the Moon was smooth and flat lasted for another two thousand years, delaying the development of Astronomy (and more precisely, its Epistemology). This leads us to observe, once again, that each era has a cut of reality, a limited capacity for cognition, which leads to a conflict between types of thoughts.

\section{Technological barriers and limitations}

In their search for more information to create greater knowledge, as discussed above, researchers need to use technological instruments, and these are not always available or developed in their time.

Seen from this perspective, technology is always a compensatory means for obtaining greater knowledge or transmitting the achieved knowledge.

Regardless of the time, the Subject-Observer tends to believe that the technology used to obtain more information about the object is always cutting-edge. This seems tautological and unreasonable, but in the 12th and 13th centuries copyists, who took around three years to copy the Bible, were considered to have cutting-edge technology. If we laugh at them today, it is out of pure cruelty. With the pace of technological advances, we will soon also be laughed at by future generations.

Another example: Portuguese caravels were also considered state-of-the-art ships. So that they would not be copied, when they lost their function, they were burned on the high seas so that they would not fall into the hands of other countries. This was the reality of the time, which did not really imagine other forms of navigation, or other means of transport and communication.

\section{Temporary epistemological limits and absolute epistemic barriers }

Given these examples and considerations, it becomes quite difficult to establish what is a temporary epistemological limit and what is an eternal epistemic barrier. An absolute epistemic barrier could, eventually be overcome with time, for example, with the invention of new technologies or approaches.

In the 17th century it was already known that human senses were not sufficient to explore the heavens. Through the telescope, in January 1610, Galileo Galilei turned the telescope to the Moon for the first time and on the following nights to Jupiter and the Pleiades. At that moment, a technological evolution began that would forever change the history of Astronomy.

Johannes Kepler had the dream of exploring the Universe, and the Earth from some other point in the Universe. This dream only became reality four centuries later.

Considering the standard model of the Big Bang, we currently do not know empirically what happened in the Universe before the era of Recombination, but this can be overcome if we resort to the detection of cosmological neutrinos and primordial gravitational waves, originating long before Recombination. The "invention" of nanotechnology by Feynman (1959), which paved the way for breaking the notion of "machine" by extending it to the microscopic world, is another additional example. Feynman recognized that there was nothing fundamental against the construction and operation of machines at a nano-scale.

In other words, there are varied examples of this type of epistemic barrier. The question is whether there are reasons to postulate that some of them are absolute, independent of time/epoch. Sometimes, there is a considerable time lag between the initial posing of the problem and its solution, if it is achieved at all. For example, the sub-Planckian regime, which describes physical entities beyond the scales governed by Planck's constant for the micro-world, the Hubble radius for the macro-world, and the event horizon of a black hole, in which assumes that the external world no longer has access to everything that goes beyond this event horizon, these are problems that have existed for several decades without definitive solutions (see below).

What is life? It's a big scientific question, and the current state of research already presents a series of discoveries that could lead to a satisfactory answer. However, it is possible that by its very nature the question of life defies a reductionist analysis, as the elements involved and their relationship may be too complex. There remains the possibility of there being an absolute epistemic barrier on this issue, and in any case the emergence of Artificial Intelligence would be a way to move towards its solution in the future. In questions of this type, the idea of nous returns with force, since we do not even know the episteme, but we certainly aim for a deep and complete knowledge of this issue that gives meaning to Humanity.

One of the characteristics of Science, of how we build it, is its permanent evolution. Therefore, it is not possible to say whether answers are definitive, they fall into the category of “René Descartes' provisional certainties”, that is, they only reflect the current stage of research and scientific consensus.

We can identify some physical and linguistic/psychological barriers that limit knowledge of the Universe (Fig.1). The very small, governed by Quantum Mechanics and with the limit of the quantum of action, the Planck constant $\hbar$ and its ``absolute'' ultimate version, the Planck scale $l_{P}$ featuring Quantum Gravity are concrete microphysical ``absolute'' scales. The very large, the size of the cosmological horizon $R_{H}$, constitute the physical extremes on the axes, while the possibility of knowing/interpreting the entire content of the Universe is limited by the language and psychology of the human being, presupposing a maximum in the field of Classical Physics. In Fig. 1 we attempted to draw a qualitative picture of the situation, as explained in the caption. However, it is important to consider these length extremes in greater detail in the current scientific context, an analysis that we will address below.

\begin{figure}[htbp]
\centering
\includegraphics[width=10cm]{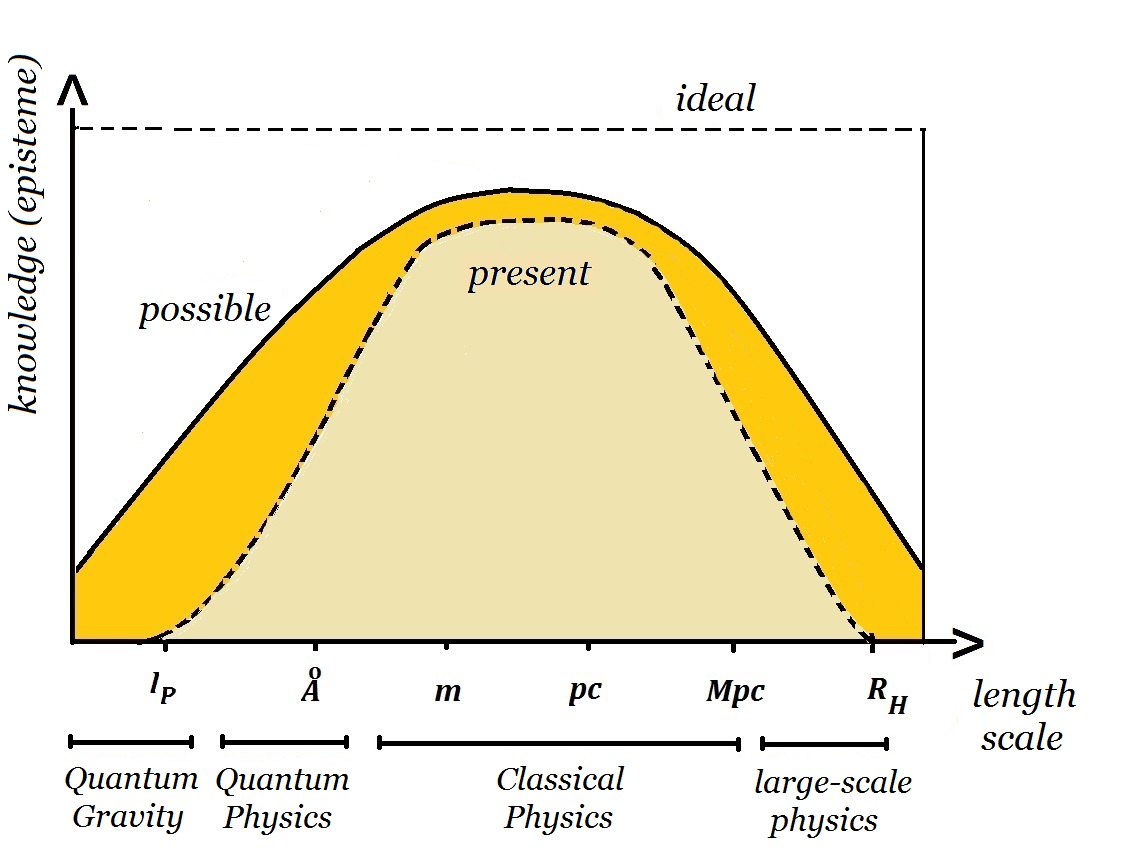}
\caption{A qualitative sketch of knowledge of Nature as a function of the scale of phenomena. Ideal knowledge (epistemic, conceptual and empirical) is represented by the dashed horizontal line at the top. Viable knowledge subject to technological, psychological and physical barriers would be achievable up to the solid black line. Current knowledge is shown by the lower dashed curved line, and varies according to the development of ideas and measurements/experiments carried out for each length scale.}
\label{fig:cint}
\end{figure}

\subsection{Quantum physics}

If asked for a description of the micro-world, any professional physicist will point to the construction of Quantum Mechanics at the beginning of the 20th century as one of the pillars of contemporary Physics. In effect, the discovery of the electron and the atomic nucleus at the beginning of the 20th century showed that the classical description, founded and elaborated for almost three centuries, and which produced Classical Mechanics, Electromagnetism and other syntheses, was not adequate to understand this new micro-world.

This is not the place to discuss the details and reasons for this rupture, but it is important to point out that the replacement of knowledge of precise values for dynamic quantities (position, velocity, etc.) by the probability of measuring one of the so-called eigenstates of the system caused a philosophical crisis that has not yet been overcome. It is well known that a group of physicists opposed this description, and insisted on the need to formulate a deterministic (not merely probabilistic) theory of the micro-world. Einstein was particularly emphatic on this, as he considered Quantum Mechanics, if not wrong, at least "incomplete" (Harrigan and Spekkens 2010). The answers from the creators of the theory, led by Niels Bohr, became increasingly strange: they declared that Reality was not the object of study of Quantum Mechanics, but only the phenomena made sense, that the measurements forced the system in question to define its state (which before these measures was undefined), and that it made no sense to talk about something that was not measured. This position has been interpreted as neo-Idealism by many philosophers, and exemplified with the famous question inspired by G. Berkeley regarding the "existence of the Moon if no one is observing it". Considering all the philosophical questions arising from this type of approach, we can say that Quantum Mechanics has qualitatively changed the type of information/knowledge that we can obtain about the micro-world: this is still knowable, but only a posteriori (after measuring and forcing the system to have a defined value of the measured quantity), and not least, accepting that there is no sharply defined ontology for microphysical objects (Bunge 2012). Clearly this must be, at some point, made compatible with classical objects, which must either be seen as Berkeley stated, or if they are finally considered "real", they will need some kind of reconciliation with the micro-world that has not yet been completely achieved (Horvath, Fernandes and Idiart 2023). It is not possible to maintain an Idealist position for microphysics and a Realist one for macroscopic objects, which are ultimately composed of the former.
An interesting development, and not much discussed, has to do with the existence of levels of elementarity proposed by David Bohm (1980). Bohm believed that there is no reason why we should stop our inquiry into the micro-world at the level of the quantum of action $\hbar$, he mentions sub-quantum scales, indicating a kind of hierarchy (could it be infinite?) towards entities of increasingly deep levels. It is unknown today whether such sub-quantum entities truly exist (Fleming 1964) and if the answer is affirmative, how can their presence be verified empirically? (Horvath 2023).

Finally, we should note that there is the possibility of resolving the paradoxes of Quantum Mechanics at the cost of admitting that quantum objects do not have defined properties for a given time, but rather a distribution of these. It should be clear that the proponents do not merely suggest a present epistemic ignorance of the Subject-Observer, but an ontology that postulates statistical values for position, momentum, etc. This hypothesis is called {\it quantons} or {\it quons} (Bunge 2012), and would force us to abandon any pretension of nous for the future. If this is the case, we would be forced to admit an uncertainty intrinsic to microphysical systems as a striking characteristic of their very existence.

\subsection{Planck's scale and Quantum Gravity}

Even deeper into the issue of elementarity, we can ask ourselves about the most extreme scale on which the description of the physical world would be possible. Twentieth-century Physics not only developed Quantum Mechanics to describe the microphysical world, but also confronted the problem of including gravitation in the description of fundamental interactions. But while Maxwell's Electromagnetism was converted into a quantum theory (bypassing serious mathematical problems, Cao and Schweber 1993), and weak and strong interactions developed in a similar way, gravitation resisted the many attempts at consistent quantization. There is a very interesting way of expressing the validity of a description where gravitation and the rest of the interactions are considered separately. The physicist Max Planck realized, at the turn of the 19th century, that there are natural units, and found a set that involves all the fundamental constants, including the then ``new'' quantum of action $\hbar$ that he introduced. Far beyond being a mere convention, Planck maintained that a system of units of this type would be useful in any physical situation, for any substance and even alien to human observers, that is, that it would define an ``absolute'' system.

The set of Planck's units is (Planck 1899)

\begin{equation}
    l_{P} = \sqrt{{\hbar G} \over{c^{3}}} = 1.6 \times 10^{-33} \, cm
\end{equation}

\begin{equation}
    m_{P} = \sqrt{{\hbar c} \over{G}} = 2.17 \times 10^{-5} \, g 
\end{equation}

\begin{equation}
    t_{P} = \sqrt{{\hbar G} \over{c^{5}}} = 5.4 \times 10^{-44} \, s
\end{equation}

\begin{equation}
    T_{P} = \sqrt{{\hbar c^{5}} \over{G k_{B}^{2}}} = 1.4 \times 10^{32} \, K
\end{equation}

which are known as Planck's length, mass, time and temperature respectively.

Although there was not much interest in Planck units initially, many decades later, as the physicists sought to reconcile theories that might span different "domains", their importance became clear. For example, Quantum Mechanics covers the micro-world, but does not describe gravitation. At the same time, General Relativity is classical, in the sense that it does not apply to the micro-world, but rather to the classical regime of macroscopic gravitating objects. Physicists realized the importance of Planck's natural units, reaching the conclusion that these values mark the limit of applicability of each domain. That is, when we try to go to smaller scales relative to Planck units, known physics is no longer valid, and more general unified theories are needed. For example, for times shorter than the Planck time $t_P$, it is not possible to consider times without a quantum theory of gravitation, and possibly the continuous time that we are familiar with needs to be replaced in this regime by a discrete time, in quantum units.

Are Planck's natural scales or units an absolute epistemological barrier? This question must be contextualized in contemporary primordial Cosmology, as there is considerable debate surrounding the so-called trans-Planckian problem, which we will describe below.

The Universe appears to correspond to flat, Euclidean geometry. There is no evidence for substantial spatial curvature. But this directly indicates, applying the existing description of General Relativity,  that the curvature in the past was still much smaller, by almost 100 orders of magnitude. Extrapolated to times $t \rightarrow 0$, the curvature would have to be truly tiny for it to be still compatible with zero today. Why is the Universe so flat? This is the so-called flatness problem. Added to this problem is the horizon problem: in very separate, even opposite, directions on the sky, the Cosmic Microwave Background is the same. But the matter that let this radiation escape had no causal contact, it is said that it was "outside the horizon". How did the Universe manage to "homogenize" the temperature in disconnected regions? This is the horizon problem.

To solve these problems, Alan Guth (1981) suggested that there was a brief stage featuring an extreme expansion of the scale factor of the Universe $a(t)$  that caused practically everything we observe today to be placed outside the horizon, and thus would explain why it became so uniform: simply this extreme expansion "inflated" a tiny piece of the Universe to gigantic dimensions, a factor $\exp(60)$ or more (Fig. 2).

\begin{figure}[htbp]
\centering
\includegraphics[width=12cm]{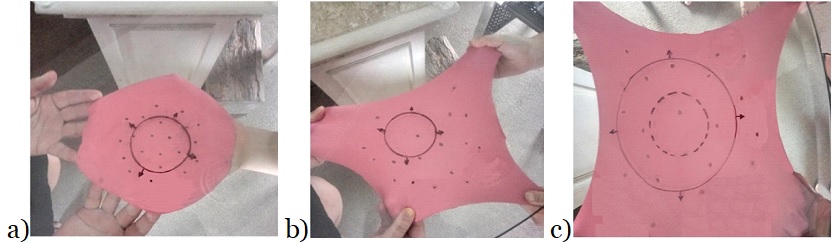}
\caption{The analogy of inflationary expansion with a stretched rubber membrane. a) A small sector of the Universe's space-time contained within the horizon (black circle) reaches conditions for $a(t)$ to grow exponentially; b) Almost all the matter originally contained within the horizon leaves it, space-time ``stretches'' and erases the curvature, producing a total flatness; c) The exponential expansion ends and the horizon advances again, progressively including an increasingly larger visible volume, while $a(t)$ continues to grow more moderately.}
\label{fig:Surv}
\end{figure}

At the highest energies that the Universe developed immediately after t=0, when its scale was on the order of the Planck length $l_P$, it is believed that it entered this inflationary phase, a stage where the scale factor of the Universe a(t)  evolves exponentially, $a(t) \propto \exp(3Ht)$, with $H = constant$ and value-dependent on the model, but it does not differ much for each of them. For now, it is not important to us what caused this Inflation. This scale factor should be thought of as the spatial substrate where matter resides. The other important quantity is the cosmological horizon, an imaginary surface that defines, for a given time, the maximum distance from which information can reach an observer. Note that this cosmological horizon, contrary to a widespread belief, is not simply c ×age, this would be the case of a Universe without expansion, for example. In the Era of Inflation, it can be shown that the horizon is stationary (does not change over time, Figure 2b), but the exponential expansion of the scale factor causes almost all matter to be placed beyond the horizon, losing causal contact. When Inflation ends, the horizon grows and progressively catches up with what previously ``left the horizon, Fig. 2c (in the case of an accelerated cosmological expansion, we will see that this is no longer true).

The interesting thing about this type of theoretical construction is that the length scales originally corresponding to a typical galaxy (now around $1 \, Mpc$), which carried the density fluctuations responsible for forming galaxies long after Inflation (Liddle 2015), were much smaller than the Planck scale of eq.(1) (Fig. 3), and for this reason we speak of a trans-Planckian problem.

Therefore, if density fluctuations were within the horizon in the Planck Era, and left it during the Inflation to finally cause the formation of galaxies, it would in principle be possible to study what happened below the Planck scale indirectly, through the statistics of galaxies that were formed, and that we see today. The Planck scale in terrestrial laboratories is practically impossible to achieve, we are today at least ten orders of magnitude below these energies, and so knowledge of trans-Planckian scales would be, if at all, empirically indirect, but certainly also accessible to the mind as Plato and Kant, for example, would agree.

\begin{figure}[htbp]
\centering
\includegraphics[width=12cm]{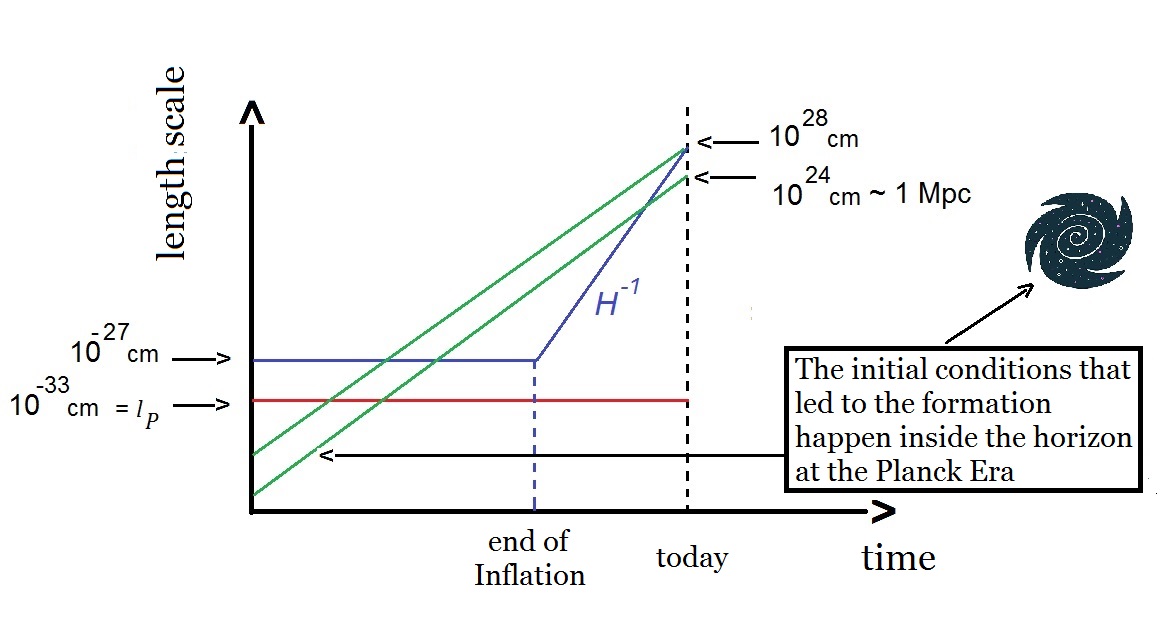}
\caption{The trans-Planckian problem. The moving scales in green ``exit'' the stationary horizon (blue horizontal line) in the Inflationary Era, and only when the cosmological horizon grows again is it able to bring these scales into the observable domain. But if this is accurate, the fluctuations that allowed galaxies to form originated at lengths smaller than the Planck length $l_P$ (beginning of the green lines, left). Thus, today we would have a way of studying this regime, at least indirectly, by studying the galaxies that were later formed.}
\label{fig:GWQ}
\end{figure}

\subsection{The accelerated Universe and cosmological horizons (large scale)}

As is the scientific consensus today, the formal Cosmology developed throughout the 20th century (known as Friedmann-Lemaître-Robertson-Walker cosmology, hereafter FLRW) indicates how to calculate the evolution of the scale factor, the cosmological horizon and all the rest of the associated quantities taking as based on Einstein's General Relativity (Landau and Rumer 2003). The contemporary novelty is that there is enormous activity to obtain a considerable amount of observational data and know the state of the Universe at all times, going well beyond a speculative mathematical description. However, these concepts are not simple to incorporate, since we are creatures used to thinking about phenomena in a fixed space-time, just as we do with mechanical collisions, wave propagation and many others.

The relationship between the increase in scale of the Universe by the Hubble expansion, 
and described quantitatively by the scale factor $a(t)$, and the cosmological horizon $R_H$ is time-varying and subtle. This immediately leads us to the next question: is there a limit to the maximum scale we can observe? Here we need to recognize that the name "horizon" is not by chance: we currently observe as far as the cosmological horizon allows, more precisely, the cosmological horizon defines the maximum distance from which we can collect information. But analogous to the ordinary horizon we see on Earth, this does not mean that there is nothing beyond this geometric limit. The problem becomes more complex when considering the fact that the horizon evolves, that it is not fixed in time, producing some persistent paradoxes. For example, there is the general idea that galaxies could only move away from us at speeds lower than the speed of light, although this is not true. The usual expression of the redshift $z$ of spectral lines in galaxies, $z=v/c$, suggests {\it prima facie} that this quantity should not exceed unity. However, this expression is only the first term of a mathematical series of the type:  $v/c+ a(v/c)^2+b(v/c)^3…$, and the first term is valid only for the closest galaxies within the Hubble flow. At the present time, the James Webb telescope has already detected very young galaxies up to $z \approx 15$, which are moving away from us at speeds much greater than that of light $c$. How is this possible? It is here that we must recognize the dynamic character of the cosmological horizon, which advances and reaches the light emitted by galaxies beyond $z \approx 1$ (Fig. 4). As a result, we observe much more of the volume of the Universe than we would if the horizon were static. This is a standard interpretation in the FLRW Cosmology, and suggests that we just need to wait long enough to see an increasingly larger volume of the Universe.

But in 1998 this interpretation suffered a setback, two independent groups announced that the Universe is expanding ever more rapidly. This acceleration of the expansion rate now changes what we thought about observing increasingly larger volumes: on the one hand, the cosmological horizon is slowing down, and simultaneously the scale factor is growing more and more quickly. Thus, something similar to the Inflation situation happens: galaxies that are close to the horizon today, leave it in the future, and we thus see less and less of the Universe. This vision is the most accepted, and predicts a dark future for observers within several Giga-years (Liddle 2015), making it philosophically interesting.

\begin{figure}[htbp]
\centering
\includegraphics[width=8cm]{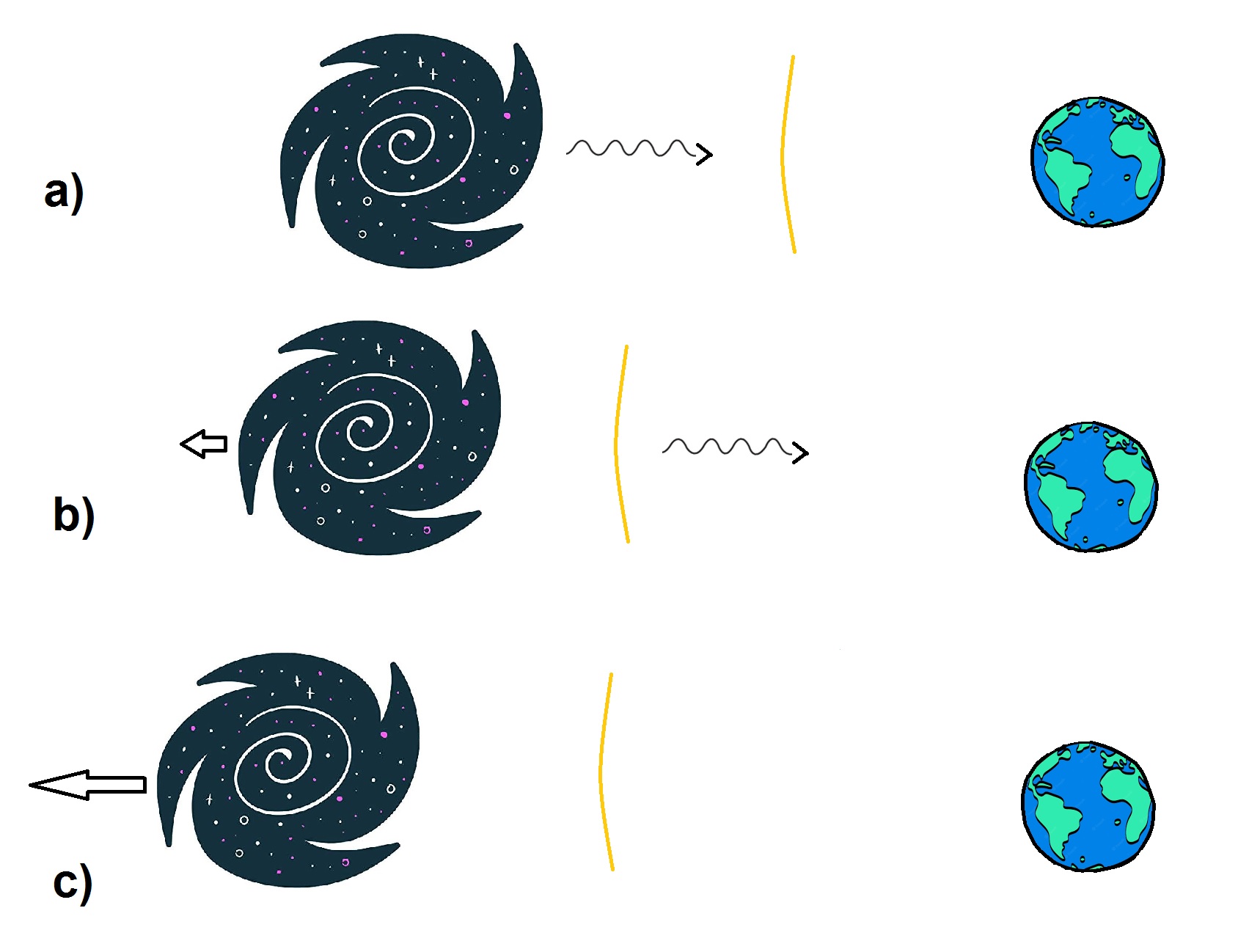}
\caption{It is possible (and in fact, it happens all the time) to see galaxies that are moving away from us at speeds much greater than the speed of light. This is possible because the Hubble expansion is not subject to the analysis of Special Relativity, being the largest self-gravitating system that exists, and whose scale factor is not a physical object. When the galaxy has already emitted light, the horizon (represented by the yellow line) advances and includes the previously emitted signals. This is why all galaxies with redshift $z \geq 1.1$ are "superluminal" and appear perfectly in the images (that is what is shown in figures a and b). However, as the current Universe is accelerating, the scale factor grows faster and faster, and thus removes galaxies from the viewing volume, thus "leaving" the cosmological horizon, as shown in figure c. This is similar to what happened in the Inflationary Era. If this acceleration persists, in the distant future we will see less and less until we lose causal contact with the rest of the Universe, eventually even with the nearby Andromeda galaxy and the galaxies of the Local Group.}
\label{fig:World}
\end{figure}

\subsection{The event horizon}

The study of solutions to Einstein's equations of General Relativity in the first half of the 20th century revealed a somewhat disturbing characteristic of a particular class: the existence of so-called event horizons. These (hyper)surfaces delimit an "inside" and an "outside" of them that causally separates the two regions. Interpreted as a central element for the concept of a black hole, they caused concern among physicists with clear philosophical connotations: if causal disconnection were taken seriously, all matter that went beyond the event horizon into the black hole would “disappear” from the black hole. instead of the accessible Universe. In other words, Nature would be able to "hide" matter and energy forever in inaccessible regions of space-time. To make matters worse, as any object in this situation carries entropy (understood here as a measure of the information contained within it), black holes would be sinks of this entropy, violating the Second Law of Thermodynamics.

With this perspective, the area that is now known as Black Hole Thermodynamics developed. The central postulate is that the area of a black hole is a measure of the entropy it carries, and thus if taken into account in the equilibrium equation, the (generalized) Second Law is not violated in spontaneous processes. At the same time, Hawking (1974) showed that black holes evaporate through the emission of quanta resulting from the quantum behavior of the vacuum around them, and thus suggested that the idea of an event horizon would be less appropriate than that of an apparent horizon (which could indeed return information to the external Universe). However, the information problem persisted, because if the escaping radiation was actually that of a black body with the temperature $T_{Hawking}$, then it would not store the information of what fell into the black hole.

However, Hawking himself, in the last years of his life, retreated from his previous positions that indicated that information was lost if matter went beyond the event horizon into a black hole, and argued that this could be recovered because the escaping radiation (called Hawking radiation) that causes it to lose mass is not totally incoherent (and thus, the phases would “remember" the state of matter or energy of the black hole, although a practical reconstruction is currently unfeasible). fundamental link between Gravitation and Quantum Mechanics, capable of clarifying the relationship between the two apparently incompatible theories. But it seems the "danger" of the definitive disappearance of matter and energy has disappeared, at least theoretically. The conditions for the evaporation of holes blacks are the ones that would allow us to recover the temporarily missing information, provided Hawking got this description right.

\section{Conclusions}

 In this article we have revisited the essential topics, starting with the limits of the senses, limitations of reason, language and epistemological methods, technology as compensation for limits, technology as the fruit of knowledge of its time, types of thoughts and the possibility of application, (at least in part, of formal, factual knowledge and the combination of both) and finally the epistemological limits and possibly absolute epistemological barriers.

While some of the discussion remained general and broad, we delved deeper into topics relating to the limits of the physical world. Our stance is that, below the Planck length $l_P$ and above the Hubble Radius scale $R_H$, it is unlikely that we will ever obtain direct information, although these domains could be indirectly accessible and analyzable by the mind (as Plato and Kant would maintain) and by observations specifics of its effects with current and future technology.

Richard Feynman argued that “we have come a long way in a short time” and that imagination necessarily leads to technological progress. From this perspective, creative capacity would have a limit within its time, space and circumstances.

It is quite likely that over time our current stage of technology will change, in one unpredictable way or another, and which will require all our knowledge to be reviewed, modified and updated. If this happens, we may reach a stage of knowledge close to absolute on all branches of knowledge, a kind of stage between a dogmatic Science –which today seems impossible– with reminiscences of the nous intended by Anaxagoras. Alternatively, some barriers may be absolute, such as the ontology of {\it quantons/quons} mentioned above, and our perception and understanding of the world should be definitively altered to accommodate a probabilistic reality.

\section{References}

\bigskip
Bohm, David. Wholeness and the Implicate Order.Routledge, NY, 1980. 

\bigskip\noindent
Bunge, Mario (2012) Does Quantum Physics Refute Realism, Materialism and Determinism? Science \& Education V. 21  (10), p. 1601-1610.

\bigskip\noindent
Cao, Tian Yu; Schweber, Silvan S. (1993) The conceptual foundations and the philosophical aspects of renormalization theory. Synthese V.97, p. 33–108.

\bigskip\noindent
Chomsky, Noam. On Language. New Press, NY, 2007.

\bigskip\noindent
Descartes, René. Discourse on Method and Meditations on First Philosophy. Hackett Publishing Company, USA, 1999.

\bigskip\noindent
Feynman, Richard Philip. Talk given at the annual American Physical Society Meeting, Caltech December 29, 1959.

\bigskip\noindent
Fleming, John J. (1964) Sub-Quantum Entities. Philosophy of Science, V. 31, No. 3, p. 271-274

\bigskip\noindent
Guth, Alan (1981) Inflationary universe: A possible solution to the horizon and flatness problems. Phys. Rev. D V.23, p.347-356

\bigskip\noindent
Harrigan, Nicholas ; Spekkens, Robert W. (2010) Einstein, incompleteness, and the epistemic view of quantum states. Found. Phys. V. 40, p.125-157 

\bigskip\noindent
Hawking, Stephen. W. (1974) Black hole explosions? Nature V. 248, p. 30-31. 

\bigskip\noindent
Horvath, Jorge E. (2023) Sub-Quantum Physics and the proton mass (submitted for publication)

\bigskip\noindent
Horvath, Jorge E.; Fernandes, Rodrigo Rosas; Idiart, Thais (2023). On the ontological ambiguity of physics facing reality. Astronomische Nachrichten V. 344, 1-2.

\bigskip\noindent
Kant, Immanuel (1999) Critique of Pure Reason (The Cambridge Edition of the Works of Immanuel Kant). Paul Guyer (Editor, Translator), Allen W. Wood (Editor, Translator). Cambridge University Press, UK.

\bigskip\noindent
Landau, Lev; Rumer, Y. What is the Theory of Relativity? . University Press of the Pacific, USA, 2003.

\bigskip\noindent
Liddle, A. An Introduction to Modern Cosmology. Wiley, NY, 2015.

\bigskip\noindent
Ortega y Gasset, Jos\'e. The Modern Theme. Forgotten Books, UK, 2018 (original 1923)

\bigskip\noindent
Planck, Max (1899) {\"U}ber irreversible Strahlungsvorgänge. Sitzungsberichte der K{\"o}niglich Preussischen Akademie der Wissenschaften zu Berlin V. 5, p. 440

\bigskip\noindent
Von Neumann, Janusz. Mathematical Foundations of Quantum Mechanics, translated by Robert T. Beyer, ed. Nicholas A. Wheeler; Princeton: Princeton University Press, 2018 (original 1932). pp. 160-164. JSTOR j.ctt1wq8zhp (1955 edition).

\bigskip\noindent
Wittgenstein, Ludwig. Tractatus Logico-Philosophicus. Routledge, NY, 2001.

\end{document}